# Analytical Study of Hexapod miRNAs using Phylogenetic Methods

A.K. Mishra and H.Chandrasekharan

Unit of Simulation & Informatics, Indian Agricultural Research Institute, New Delhi, India

akmishra@iari.res.in , akmishra.usi.iari@gmail.com

## ABSTRACT

MicroRNAs (miRNAs) are a class of non-coding RNAs that regulate gene expression. Identification of total number of miRNAs even in completely sequenced organisms is still an open problem. However, researchers have been using techniques that can predict limited number of miRNA in an organism. In this paper, we have used homology based approach for comparative analysis of miRNA of hexapoda group .We have used Apis mellifera, Bombyx mori, Anopholes gambiae and Drosophila melanogaster miRNA datasets from miRBase repository. We have done pair wise as well as multiple alignments for the available miRNAs in the repository to identify and analyse conserved regions among related species. Unfortunately, to the best of our knowledge, miRNA related literature does not provide in depth analysis of hexapods. We have made an attempt to derive the commonality among the miRNAs and to identify the conserved regions which are still not available in miRNA repositories. The results are good approximation with a small number of mismatches. However, they are encouraging and may facilitate miRNA biogenesis for hexapods.

## KEYWORDS

miRNA, homology, alignment, conservation, sequences

## 1. INTRODUCTION

miRNAs are an evolutionary conserved class of non coding RNAs of small length approximately 20 to 25 nucleotides (nt) long and found in diverse organisms like animal, plant etc. miRNAs play very important role in various biological processes. They regulate gene expression at post transcriptional level by repressing or inactivating target genes [1, 2]. miRNA biogenesis is highly associated with stem-loop feature of its precursor's secondary structure. As pre-miRNA secondary structures consisting of stem-loop are highly conserved across different species, extracting informative attributes from secondary structure is significant step in identification of miRNA from unknown sequences [3]. The biochemical based methodology used for identification of novel miRNAs, in the laboratories can be assisted by computational methods. Computational methods can identify various potential miRNA that can further be verified by the former approach. Therefore, researchers have developed computational models for miRNA prediction. miRNA gene finding is a challenging research area in the field of computational biology. This is evident from various studies that hairpin structure of miRNA is dominating and accounts for its secondary structure and quite informative in retrieving and inferring biological information [4].

Since eukaryotic genome contains high number of inverted repeats, which are on transcription, converts into hairpin structure. Due to its large number in the genome it has become quite important to choose the right hairpin among them and this is a major problem that biologists suffer. Computational methods have given an attempt to reduce the search space, which therefore proved helpful in searching the right hairpin for prediction. There are different approaches to predict the secondary structure of RNA. These are classified as energy minimization based, grammar based, matching based and evolutionary algorithm based approaches. Free energy minimization is one of the most popular methods for the prediction of secondary structure of RNA. Energy minimization methods use the dynamic programming approach along with some sophisticated energy rules. Energy of the predicted RNA structure is estimated by summing negative base stack energy of each base pair and by adding positive energy of destabilizing regions like loops, bulges and other unpaired regions[5..8] .

Three fundamental techniques used to isolate miRNA are forward genetic screening, direct cloning, and computational analysis through bioinformatics tools. But the most common method is, to isolate and clone the miRNA obtained though biological samples and this method has been widely used in plant miRNA identification. However, miRNAs are difficult to clone due to its short length, very low expression level at specific cell and under specific condition. Therefore, it is clear that computational techniques are needed for identification of miRNA genes in any sequenced genome. The same is true for the homology based search for which there was no clear evolutionary model [9, 10]. Therefore, a series of computational tools has been developed for the computational prediction of miRNA and to find out the homologue between the species and sub species. In homology based models miRNA of known species is taken as a query sequence and the same is searched in another species [11, 12].

## 2. MOTIVATION

The present scenario is gaining much preference in the computational identification of non-coding region of RNA. However, it is difficult to study the non-coding regions of RNA as compared to the coding region of RNA because it does not possess a strong statistical signal as well as it lacks generalized algorithm. miRNA and pre-miRNA prediction is still not a widely explored research area in the domain of computational biology. Identification of total number of miRNAs even in completely sequenced organisms is still an open problem. However, researchers have been using limited techniques for miRNA prediction.

Further, even most of these tools and techniques are used for miRNA prediction of animals and plants genome. To the best of our knowledge very little information is available to hexapoda group miRNA analysis. It is always fascinating when a problem has been analysed for a long period of time and no clear cut methodology for optimal solution are developed. miRNA analysis of hexapoda group also falls in this category of problem. This forms the motivation for the present work.

## 3. REVIEW OF HOMOLOGY BASED APPROACHES

Previously, miRNA gene searches were carried out on the basis of both sequence and structure conservation between two closely related species. Now it is known that for a suitable candidate hairpin region should be more conserved than pre-miRNA along with certain other criteria [13].

Approaches for computational prediction of miRNA can be categorized as: Filter based, Machine learning, Target centered, Homology based, Rule based and Mixed. Most of the prediction methods uses homology-based search as a major part of their protocol and even conservation requirement is integral part in orthologues search. There are many homology based approaches which are applicable to the original candidates set that usually fail to pass the filters. These methods are very useful for finding new members of conserved gene clusters of miRNAs. They are useful in finding newly sequenced genomes for homologues of known miRNAs and for validation of miRNA genes in the previously known genomes [14..16]. However, these methods can be refined by incorporating structure conservation. ERPIN is a profile-based method [17] which uses both sequence and structure conservation and has successfully predicted hundreds of new candidates belonging to different species. Another similar approach is miRAlign [18], which has been proved successful over the conventional search tools.

## 4. MATERIALS AND METHODS

The work reported in this paper has been carried out in following four phases:
- Data collection
- Determination of common mature sequences
- Pair wise alignment
- Multiple sequence alignment

### 4.1 MicroRNA Data Collection

There are very limited open and free domain sources of miRNA data available to the research community. miRBase is one of the highly referred database easily accessible for miRNA research and in latest release 10883 pre- miRNAs are available. We downloaded a dataset of 62 known pre-miRNA of Apis mellifera, 66 from Anopholes gambiae, 91 from Bombyx mori and 157 from Drosophila melanogaster respectively from miRBase sequence database (a data repository of published microRNA sequences and its annotation) (release 14.0) at http://microrna.sanger.ac.uk [19,20].

### 4.2 Determination of common mature miRNAs among species

A perl script is written to analyse the miRNA similarity among different hexapoda organisms. We have derived similarities in group of 2, 3 and 4 species. The detail results with all combinations are presented using van diagram and discussed in the next section.

### 4.3 Pair wise alignment using Blast

In this phase, known miRNA of Bombyx mori is used as query sequence for similarity search against complete genomic sequence of Apis melleifera and Droshophila melanogaster to generate a set of candidate homologues. We used BLAST [21] with E-value cutoff of 10 and minimal word size of 7 since sequence is very much conserved across large evolutionary distances. We have generated hits for almost all miRNA homologues. During filtering step, we discarded sequences of the candidate set in which alignment region does not fall in the mature part of the query sequence. We also discarded candidates in which length of the mature sequences differs by more than 2 nt.

### 4.4 Multiple sequence alignment using Clustral W

In this phase to obtain families of homologous miRNAs, we started with a CLUSTAL-W [22] alignment of all miRNA precursors in the miRBase repository. From a visual inspection of the CLUSTAL-W generated trees, we extracted miRNA clusters, ranging in size from 8 to 27 sequences. Precursor sequences contained within each cluster were not necessarily homologous, but they were close enough in terms of sequence and structure to produce useful hints for subsequent search.

## 5. RESULT AND DISCUSSION

We have done pair wise alignment of all 91 miRNAs with the complete genomic sequence of Drosophila melanogaster and Apis mellifera with E value to 10 and word size of 7 to generate homologues sequences using BLAST. Some examples are mentioned below.
Bombyx mori miR-1 (uggaauguaaagaaguauggag) vs. Drosophila genome

```
Query  1         TGGAATGTAAAGAAGTATGGAG  22
                 ||||||||||||||||||||||
Sbjct  20487496  TGGAATGTAAAGAAGTATGGAG  20487517
```
Bombyx mori let-7 (ugagguaguagguuguauagu) vs. apis genome
```
Query  1      TGAGGTAGTAGGTTGTATAGT  21
              |||||||||||||||||||||
Sbjct  75213  TGAGGTAGTAGGTTGTATAGT  75233
```
In both the results similarity found is 100 %. In the following table Table1 the result of Bombyx mori miRNA is shown against Apis and Drosophila genomic sequences considering mismatches up to 2.

**Table 1.**  (Pair wise alignment of Bombyx mori, Drosophila melanogaster and Apis mellifera)

| Species vs. Species | 100 % match | One mismatch | 2 mismatch | Not significant |
|---|---|---|---|---|
| Bombyx vs. Drosophila | 66 | 20 | 4 | 1 |
| Bombyx vs. Apis | 62 | 19 | 6 | 4 |

We have derived similarity in group of 2, 3 and 4 species. The results for intersection of all combinations between two organisms are given in table2 and for grouping of 3 and 4 species are presented using van diagram as mentioned in figure 1.

Table 2. (No. of miRNA matches between species)

| Species Vs Species matches | Drosophila melanogaster | Bombyx mori | Apis mellifera | Anopheles gambiae |
|---|---|---|---|---|
| Drosophila melanogaster | X | 44 | 48 | 53 |
| Bombyx mori | 44 | X | 36 | 38 |
| Apis mellifera | 48 | 36 | X | 46 |
| Anopheles gambiae | 53 | 38 | 46 | X |

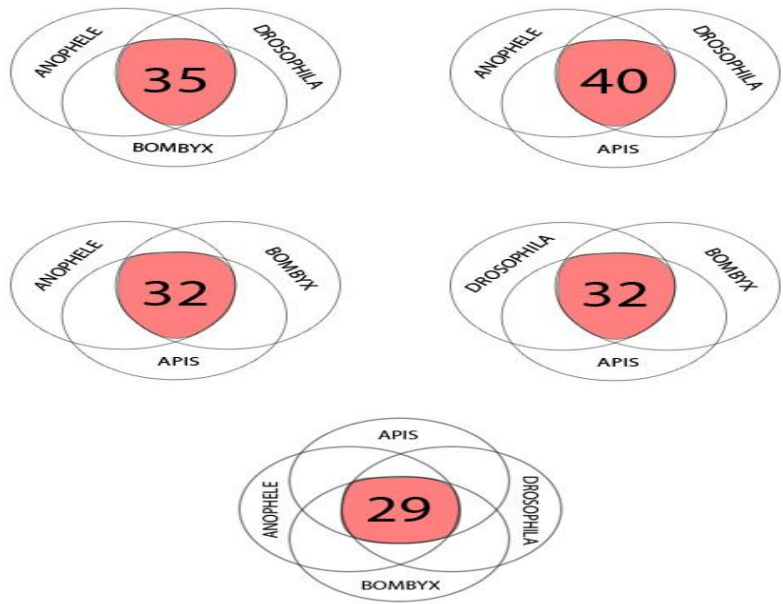

Fig. 1. (Van diagram to represent intersection among hexapoda species)

We started with a CLUSTAL-W alignment of all miRNA precursors in the miRBase for hexapoda group. From a visual inspection of the CLUSTAL generated trees, we extracted miRNA clusters, ranging in size from 8 to 27 sequences. During the analysis of the results for conservation among Apis mellifera, Bombyx mori and Anopheles gambiae we found that approximately 82 % mature sequences fall under conserved region, and 13 % mature sequences are outside the conserved region by 1-2 nt.

An exceptional miRNA (mir-276) is detected with two conservation regions. Apis mellifera mature sequence is detected in one region while Bombyx mori mature sequence in the other conserved region as shown below.

**mir-276**

CLUSTAL 2.0.10 multiple sequence alignment
```
aga-mir-276        --GGU-GACUGCCAUCAGCGAGGUAUAGAGUUCCUACGGUAAUCGAUUGAAACUUUGUAG 57
ame-mir-276        -UGGUAGAGAUCCAGCAGCGAGGUAUAGAGUUCCUACG-UAGUGUUCAGAAA----GUAG 54
bmo-mir-276        CUGGU-AAUUACCACUAGCGAGGUAUAGAGUUCCUACG----UAUGCUAACACU--GUAG 53
                        ***   *   ***  ********************     *    *  *    ****

aga-mir-276        GAACUUCAUACCGUGCUCUUGGA-UAGCCGUUUACC 92
ame-mir-276        GAACUUCAUACCGUGCUCUUGGACUUGCCG------ 84
bmo-mir-276        GAACUUCAUACCGUGCUCUUGGGUUUGCCAA----- 84
                   **********************  *  ***
```
Conserved region in the sequence
-> Position of 1$^{st}$ conserved region = 25nt
```
aga-mir-276        CCAUCAGCGAGGUAUAGAGUUCCUACG
ame-mir-276        CCAGCAGCGAGGUAUAGAGUUCCUACG
bmo-mir-276        CCACUAGCGAGGUAUAGAGUUCCUACG
```
(contain bmo mature seq)

-> Position of 2$^{nd}$ conserved region = 26nt
```
aga-mir-276        GUAGGAACUUCAUACCGUGCUCUUGG
ame-mir-276        GUAGGAACUUCAUACCGUGCUCUUGG
bmo-mir-276        GUAGGAACUUCAUACCGUGCUCUUGG
```
(contain ame mature seq)

# 6. CONCLUSION

Conservation analysis carried out for Apis mellifera, Bombyx mori, Drosophila melanogaster and Anopheles gambiae in this work revealed that theses four species are interrelated. Conservation is proved to be suitable criteria to further study the relationship and biogenesis among these species. Conservation region in alignments showed that the mature sequence comes exactly under it. While carrying out the similarity search we found that some alignments exhibited two regions of conservation wherein first region has the mature sequence but the functionality for second is unknown and it is not exactly falling in the complementary region of the mature miRNA. These results show that the other region which does not have mature sequence in it can have the full potential to be a mature region. So this finding can be further validated using some wet lab experiments to determine its potentiality. However, sequence alignment alone may fail to identify miRNAs that diverged too far. We expect that a more optimal approach exploiting information from both miRNA structure and sequence could significantly improve precision and recall of homologous miiRNA genes.

# REFERENCES
[1] Lee RC, Feinbaum RL, Ambros V (1993) The *C. elegans* heterochronic gene *lin-4* encodes small RNAs with antisense complementarity to *lin-14*. *Cell* **75**: 843-854
[2] Lewis BP, Shih IH, Jones-Rhoades MW, Bartel DP, Burge CB (2003) Prediction of mammalian microRNA targets. *Cell* **115:** 787-798
[3] Bartel,D.P. :MicroRNAs :Genomics, biogenesis, mechanism and function. *Cell* 116 (2004) 281–297
[4] Lee,Y. et al. The nuclear RNaseIII Drosha initiates microRNA processing *Nature* 425(2003)415–419
[5] Tinoco Jr., I., Uhlenbeck, O.C., and Levine, M.D. (1971): Estimation of secondary structure in ribonucleic acids. *Nature* 230: 362–367
[6] Zuker, M. and Stiegler, P. (1981): Optimal computer folding of large RNA sequences using thermodynamics and auxiliary information. *Nucleic Acids Res.* 9: 133–148.
[7] Nussinov, R. and Jacobson, A.B. (1980): Fast algorithm for predicting the secondary structure of single-stranded RNA. *Proc. Natl. Acad. Sci.* 77: 6309–6313.
[8] Mathews, D.H., Sabina, J., Zuker, M., and Turner, D.H. (1999): Expanded sequence dependence of thermodynamic parameters improves prediction of RNA secondary structure. *J. Mol. Biol.* 288: 911–940.
[9] Mette MF, van der Winden J,et al (2002). Short RNAs can identify new candidate transposable element families in Arabidopsis. *Plant Physiol.* 30:6–06
[10] Sunkar R, Girke T, et al (2005) Cloning and characterization of MicroRNAs from rice. *Plant Cell* 17:1397–411
[11] Xie Z, Kasschau KD, Carrington JC. (2003) Negative feedback regulation of Dicer-Like1in Arabidopsis by microRNA-guided mRNA degradation. *Curr. Biol.* 13:784–89
[12] Jones-Rhoades MW, Bartel DP. (2004) Computational identification of plant MicroRNAs and their targets including a stress-induced miRNA. *Mol. Cell* 14:787–99
[13] Mendes N D, Feeitas A T, Sagot M F (2009) Current tools for the identification of miRNA genes and their target *Nucleic Acid Research* 1-15
[14] Xie x, Lu J Kulbakas E J, Golub TR, et al (2005) Systematic discovery of regulatory motif in human promoters and 3' UTR by comparision of several mammals *Nature* 434:338-345
[15] Chatterjee,R. and Chaudhuri,K. (2006) An approach for the identification of microRNA with an application to Anopheles gambiae. *Acta Biochim. Pol.,* 53: 303–309.
[16] Weaver,D., Anzola,J., et al (2007) Computational and transcriptional evidence for microRNAs in the honey bee genome. *Genome Biol.*, 8: R97.
[17] Legendre,M., Lambert,A. and Gautheret,D. (2005) Profile-based detection of microRNA precursors in animal genomes *Bioinformatics*, 21:841–845.
[18] Wang,X., Zhang,J., et al (2005) MicroRNA identification based on sequence and structure alignment. *Bioinformatics,* 21:3610–3614
[19] Griffiths-Jones S, Saini Hn Dongen S, Enright AJ miRBase tools for microRNA genomics *NAR 2008 36(database issue)* D154-D-158
[20] Griffiths-Jones S, Grocock RJ, Van Dongen SBatchan A, Enright AJ miRBase: microRNA 2006 sequences and gene nomenclature *34(Database issue)* D140-D144

[21] Stephen F. Altschul, Warren Gish, Webb Miller Eugene W. Myers and David J. Lipmanl Basic Local Alignment Search Tool *J. Mol Biol* (1990) 215,403-410

[22] JD Thompson, TJ Gibson, F Plewniak, F Jeanmougin and DG Higgins The CLUSTAL_X windows interface: flexible strategies for multiple sequence alignment aided by quality analysis tools *Nucleic Acids Research*, Vol 25, Issue 24 4876-4882